\documentclass[useAMS,usenatbib]{mn2e}
\usepackage{graphicx}
\usepackage{lscape}
\usepackage{longtable}
\usepackage{rotating}
\usepackage{colordvi}
\usepackage{color}
\usepackage[english]{babel}
\usepackage{amsmath}
\usepackage{amssymb}
\usepackage{amsfonts}
\usepackage{epsfig}
\usepackage{subfig}
\usepackage{natbib}

\title[MOND and Dark Matter Halo for  ESO138-G014]{Rotation Curve with MOND and Dark Matter Halo profile for ESO138-G014}
\author[ Hashim, De Laurentis, Zainal Abidin, Salucci,]{Norsiah Hashim$^{1}$, Mariafelicia De Laurentis$^{2,3,4}$\thanks{E-mail:
mfdelaurentis@tspu.edu.ru; }, Zamri Zainal Abidin$^{1}$, Paolo Salucci$^{5}$\\
$^{1}$Radio Cosmology Research Lab,Physics Dept., Faculty of Science,
	University of Malaya, Kuala Lumpur,Malaysia\\
$^{2}$Tomsk State Pedagogical University, 634061 Tomsk and National Research Tomsk State University, 634050 Tomsk, Russia\\
$^{3}$Dipartimento di Fisica, Universit\`{a} di Napoli {} "Federico II'', Compl.
Univ. di Monte S. Angelo, Ed. G, Via Cinthia, I-80126, Napoli, Italy\\
$^{4}$INFN Sezione di Napoli, Compl. Univ. di Monte S. Angelo, Edificio G, Via
Cinthia, I-80126, Napoli, Italy\\
$^{5}$Astrophysics Sector, SISSA/ISAS, International School for Advanced Studies,
        	Via Bonomea 265, 34136, Trieste,Italy	}
\begin{document}


\pagerange{\pageref{firstpage}--\pageref{lastpage}} \pubyear{2014}

\maketitle

\label{firstpage}

\begin{abstract}
This paper is devoted to solve the galactic rotation problem for ESO138-G014 galaxy based on two theories: dark matter and Modified Newtonian Dynamics. Here we did the rotation curve analysis with two possible choices for the dark matter density profile, namely Burkert and Navarro, Frenk and White profiles. The analysis shows the dark matter distribution favored to Burkert profile (cored dark matter). The standard hypothesis for most spiral galaxies are known to be embedded in dark matter haloes has now been overshadowed by Modified Newtonian Dynamics, known as MOND, the leading alternative of dark matter. MOND addresses the problem of a new fundamental constant $a_{0}$, called the acceleration constant, at which acceleration scale of Newtonâ second law fails to hold. In this respect, we investigate this issue by testing the rotation curve within the MOND framework with the observations to obtain the reliable disk mass, $M_D$. We investigate whether ESO138-G014 is compatible with MOND or dark matter is still favorable for the galactic rotation problem.\

\end{abstract}


\section{Introduction}
The missing mass controversy begun in the $1930$â when \cite{Oort1,Oort2} and \cite{Zwicky} independently found the evidence for the vast amount of unseen matter in different scales~\cite{O'Brien:10}. Then in $1970$'s, the modern dark matter research began with many papers that found that galaxies contain more gravitating matter which will be accounted for by the stars only. The pioneering work by \cite{rubin:80} and \cite{bosma:81} found that the required mass of many spiral galaxies is much larger than the observed mass of all the visible stars and gas. Eventually, despite the widely acceptance of dark matter theory by the scientific community, an attempt to omit dark matter as the solution for the rotation curve problem has been done in the past decades by modifying the gravitational law. The most interesting, yet the one and only attempt that survived the observational tests is the Modified Newtonian Dynamics (MOND), advocated by 
\cite{milgrom:08}, \cite{sanders:07} and \cite{de:98}. By considering the mass-to-light ratio of the stellar disk as free parameter, MOND succesfully reproduced the observed rotation curve without any dark matter (see \cite{sanders:02, milgrom:08, sanders:07, swaters:10}). The success of the MOND theory as an alternative solution of the missing mass problem in extragalactic astronomy leads us to the study of the rotation curve of the ESO$138$-G$014$ galaxy within the MOND procedure.

If the MOND theory is supposedly ruled out in the ESO$138$-G$014$ galaxy, then dark matter will be a sufficient solution for the galactic rotation problem. The dark halo models that we studied in this paper are cored or cuspy halo. It is commonly referred as the core-cusp problem. We test which model is consistent with the observation of galactic halo.


ESO$138$-G$014$ galaxy, was observed at distance $d=18.57$ Mpc with the Australia Telescope Compact Array (ATCA) on $8-9$ Dec and $29-30$ Nov in $2002$. This object is a result of our screening selection from the literature from year $2009$ till $2011$  by \cite{O'Brien:10}. 
The object was selected due to several criterias: 
\begin{enumerate}
\item good quality of $21$-cm rotation curve, 
\item distance, known with precision, in that the distance is such that redshift is a good distance indicator
\item no significant existence of bulge,
\item stellar photometry and 
\item HI surface density.
\end{enumerate}
 ESO$138$-G$014$ was first studied in HI in $1982$ with the Las Campanas $1-m$ telescope by \cite{aaronson:82}. Note that, this galaxy is a relatively large spiral galaxy of the type Sd that with HI disk extended to $\sim 21$ kpc and $110.97$ km s$^{-1}$ maximum rotation speed. Nevertheless,~\cite{O'Brien:10} reported that they missed the extended structure on large spatial scales due to the lack of short spacing during observation.
 
 {\bf Outline} : The remainder of this article is organized as follows.
In Section~\ref{RCAnalysis} gives account of rotation curve analysis for ESO$138$-G$014$ galaxy.
In Section~\ref{MOND}, we look at how rotation curve in MOND framework will work on the observed rotation curve. 
Finally, we discuss our results and conclude in Section~\ref{conclusions}.

\section{Rotation Curve Analysis}\label{RCAnalysis}
Rotation curve analysis is one of the overwhelming evidence for the dark matter existence in galaxies. It is a fair measure of its gravitational potential that make up the dynamical of the galaxies in equilibrium state. Generally, a galaxy is made of several components, stars, gas and also dark matter. The ratio of these three components varies from one galaxy to another. The study of the velocity field of stars and gas is crucial in order to map out the mass distribution of galaxies. This section will emphasize on rotation curve modeling for ESO$138$-G$014$ galaxy.

\subsection{HI Mass}\label{HI}
The total HI mass, $M_{HI}$ is measured by the formula as follow:
\begin{equation} 
M_{HI}\left( r\right) =2\pi \int_{0}^{r}r\sigma \left( r\right) dr, \label{eqMHI}
\end{equation}
where $\sigma(r)$ is the HI surface density given by~\cite{O'Brien:10}. The total HI mass calculated from equation (\ref{eqMHI}) yields: $M_{HI}\sim 6.3\times 10^{9}M_{\odot }$ within $r\sim 20$ kpc. However, as we approximate the HI mass from the real data of $\sigma(r)$, and with the same equation (\ref{eqMHI}), we obtained $M_{HI}\sim 6.4\times 10^{9}M_{\odot }$ within $20.6$ kpc. Please refer to Figure~\ref{fig:HImass} to complement and clarify the discussions that follow. The profile of HI mass in Figure~\ref{fig:HImass} will subsequently gives the estimation of rotational velocity for gas contribution. A rough estimation of $V_{gas} (r)$ can be made by considering, $V_{gas}(r)=\sqrt{\frac{GM_{HI}\left( r\right) }{r}}$. 
\begin{figure}
\includegraphics[scale=0.7]{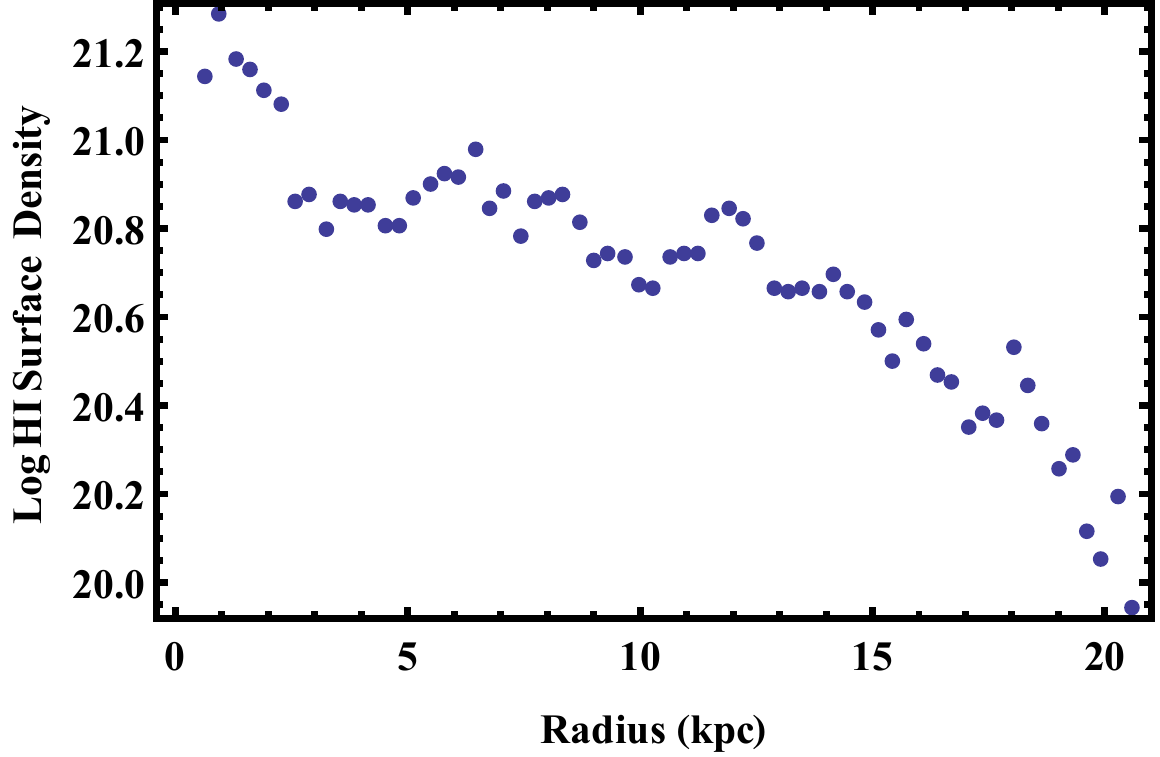}
\caption{The real data of HI surface density $\sigma(r)$, as given by van der Kruit.}
\label{fig:HIsurfaceDensityReal}
\end{figure}
\begin{figure}
\includegraphics[scale=0.7]{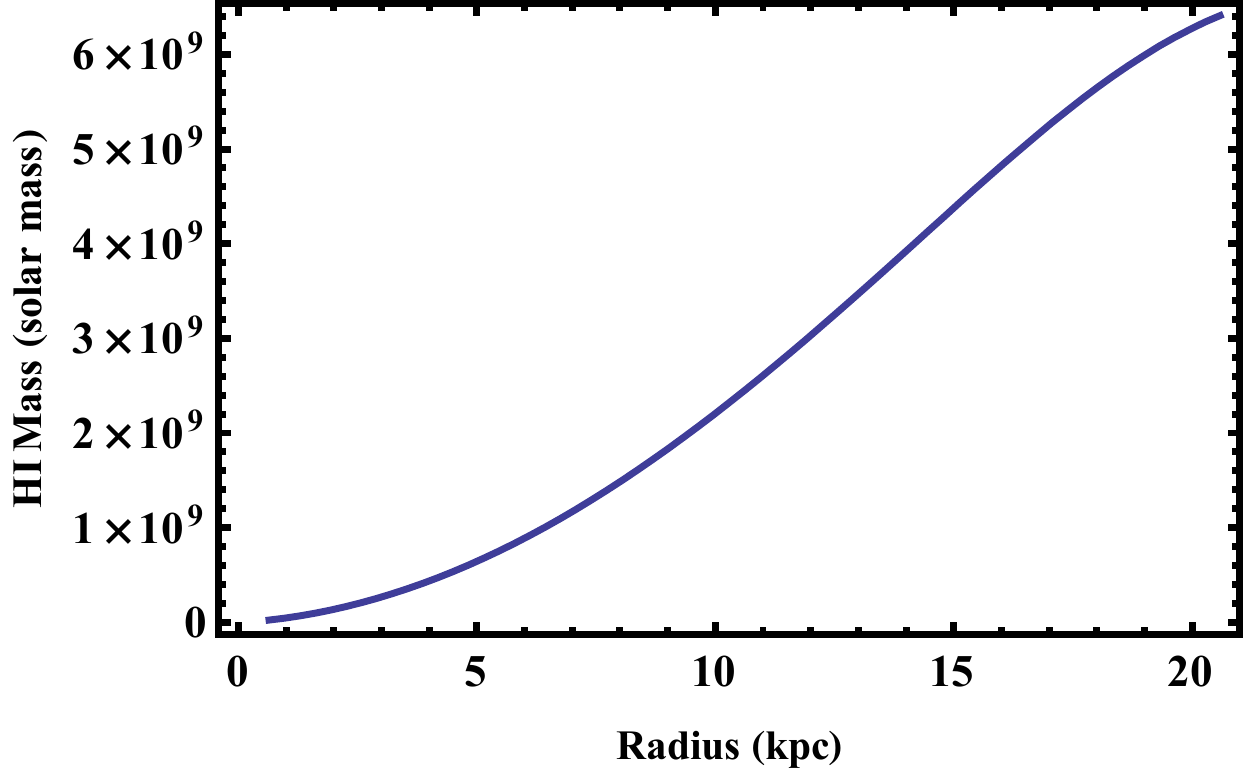}
\caption{The derivation of neutral hydrogen mass from equation (\ref{eqMHI}) with the HI surface density $\sigma(r)$, in Figure~\ref{fig:HIsurfaceDensityReal}.}
\label{fig:HImass}
\end{figure}
\subsection{Stellar Mass: Surface Brightness}\label{Stellar}
\begin{figure}
\includegraphics[scale=0.7]{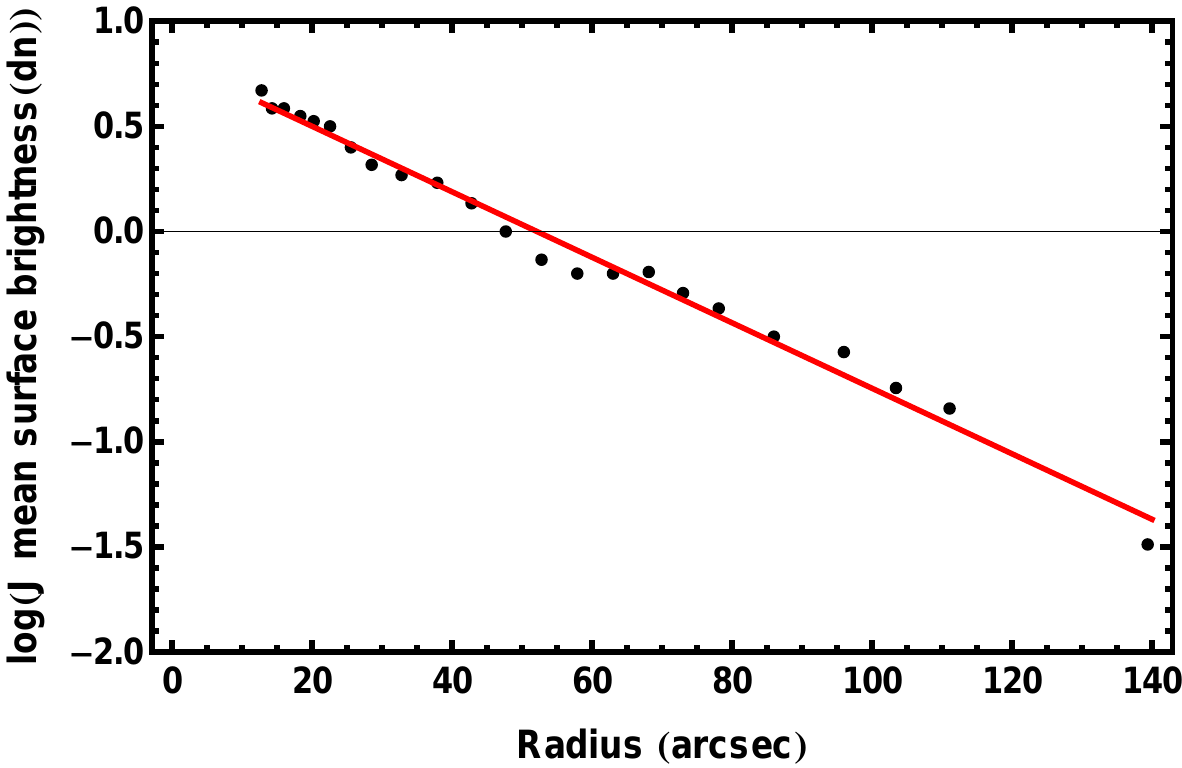}
\caption{The surface brightness profile of the ESO$138$-G$014$ in J-band with the linear fitting function.}
\label{fig:LinearFitting}
\end{figure}

The second information that we need for our task is to obtain the approximation of the disk scale length from the surface brightness profile of the galaxy. As we all know, the surface brightness profile of the disks has been proposed to have an exponential fall-off that corresponds to the equation below,
\begin{equation}
\Sigma(r)=\Sigma_{0}e^{-\frac{r}{R_{D}}}, \label{surfBrightness}
\end{equation}
where $r$ is the 2-dimensional radial coordinate in the plane of the disk and $\Sigma_{0}$ and $R_D$ are constants. Following the equation~(\ref{surfBrightness}) and solved simultaneously with the fitting function of the surface brightness profile to get $R_D$, the disk scale length. The surface photometry of ESO$138$-G$014$ was taken from the 2MASS (Two Micron All Sky Survey) database to derive the luminosity profile. The plot of the fitting is presented in Figure~\ref{fig:LinearFitting}.
The intersection between equation (\ref{surfBrightness}) and linear fitting function gives the value of the disk scale length, $R_{D}\sim\,30$ arcsec or $\approx 2.7$ kpc at the adopted distance from \cite{O'Brien:10}. Throughout this paper, we will use this value for our modeling.

\subsection{Kinematics of ESO138-G014}\label{RC}
While all the luminous component for the object has been studied, the presence of the dark component (dark matter) will now be sufficient in our modeling. This simple model can be viewed as a sum of the disk, halo and gas components as reported in the next subsubsection.
\subsubsection{Rotation Curve Decomposition}\label{RC1}
Many authors emphasized that mass modeling is uncertain (see \cite{van:85}, \cite{skillman:87} and \cite{lake:89}). In this paper, we will decompose the mass modeling to a multicomponent model as generally, visible matter more dominating at the central part of the rotation curve. Therefore, we consider the mass model comprising disk, halo and gas component,
\begin{equation}
V_{total}^{2}=V_{disk}^{2}+V_{halo}^{2}+V_{gas}^{2},\label{Vtotal}
\end{equation}
where, $V_{gas}$ is derived from Figure~\ref{fig:HImass} and $V_{disk}^2$ as in equation below: 
\begin{equation}
\bigskip V_{disk}=\sqrt{\frac{0.5GM_{D}\left( 3.2x\right) ^{2}\left(
I_{0}K_{0}-I_{1}K_{1}\right) }{R_{D}}}, \label{Vdisk}
\end{equation}
is a rotation velocity from disk contribution of a galaxy. The modified Bessel functions are computed at $1.6x$, with $x=\frac{r}{R_{opt}}$  ($R_{opt}=3.2R_{D}$), $R_{D}\thickapprox 30$ arc sec and the disk mass, $M_{D}$ is a free parameter. While for the dark halo component, we will consider cored and cuspy model and compared with observational data from ATCA. 
\cite{burkert:95} and \cite{salucci:03a}) defined dark halo component as follows,
\begin{eqnarray}
V_{halo}^{2}&=&6.4G\frac{\rho _{0}r_{0}^{3}}{r}\left\{ \ln \left( 1+\frac{r%
}{r_{0}}\right) -\arctan \left( \frac{r}{r_{0}}\right) +\right.\nonumber\\&& +\left.\frac{1}{2}\ln
\left[ 1+\left( \frac{r}{r_{0}}\right) ^{2}\right] \right\}, \label{VhaloBurkert}
\end{eqnarray}
with density,
\begin{equation}
\rho_{b} \left( r\right) =\frac{\rho _{0}r_{0}^{3}}{\left( r+r_{0}\right)
\left( r^{2}+r_{0}^{2}\right)}\,,\label{Burkdensity}
\end{equation}
where it is parameterized by $\rho_{0}$ and  $r_{0}$ are, respectively, the central density and core radius. Then the corresponding mass is,
\begin{eqnarray}
M_{b}(r)&=& 6.4\,\rho _{0}\,r_{0}^{3}\left\{ \ln \left( 1+
\frac{r}{r_{0}}\right) -\arctan \left( \frac{r}{r_{0}}\right) +\right.\nonumber\\&& +\left.\frac{1}{2}
\ln \left( 1+\left( \frac{r}{r_{0}}\right) ^{2}\right) \right\}. \label{Mburkert}
\end{eqnarray}
Likewise, we also have chosen another dark halo model, the NFW profile, which is a simple formula proposed by Navarro, Frenk and White (see~\cite{navarro:97} and \cite{power:03}) as the universal profile resulting from predictions of standard cold dark matter (CDM) cosmology $N$-body simulations, and is extensively used in the literature. They also predicted the extended dark matter halo around galaxies. The corresponding profile is given as follows
\begin{equation}
\rho _{NFW}\left( r\right) =\frac{\rho _{0}}{\left( \frac{r}{r_{s}}\right)
\left( 1+\frac{r}{r_{s}}\right) ^{2}}, \label{NFWdensity}
\end{equation}
with scale radius, $r_{s}$. The profile describes $\rho \left( r\right) \propto r^{-1}$ for $r\ll r_{s}$ and $\rho \left( r\right) \propto r^{-3}$ for $r\gg r_{s}$. The velocity contribution of this profile, is as below:
\begin{equation}
V_{halo}^{2}=\frac{4\pi G\rho _{0}r_{s}^{3}}{r}\left\{ \ln \left( 1+\frac{r%
}{r_{s}}\right) -\frac{\frac{r}{r_{s}}}{\left( \frac{r}{r_{s}}\right) +1}%
\right\}\,. \label{VhaloNFW}
\end{equation}
\subsubsection{Data Fitting}\label{Fit}
\begin{figure}
\includegraphics[scale=0.55]{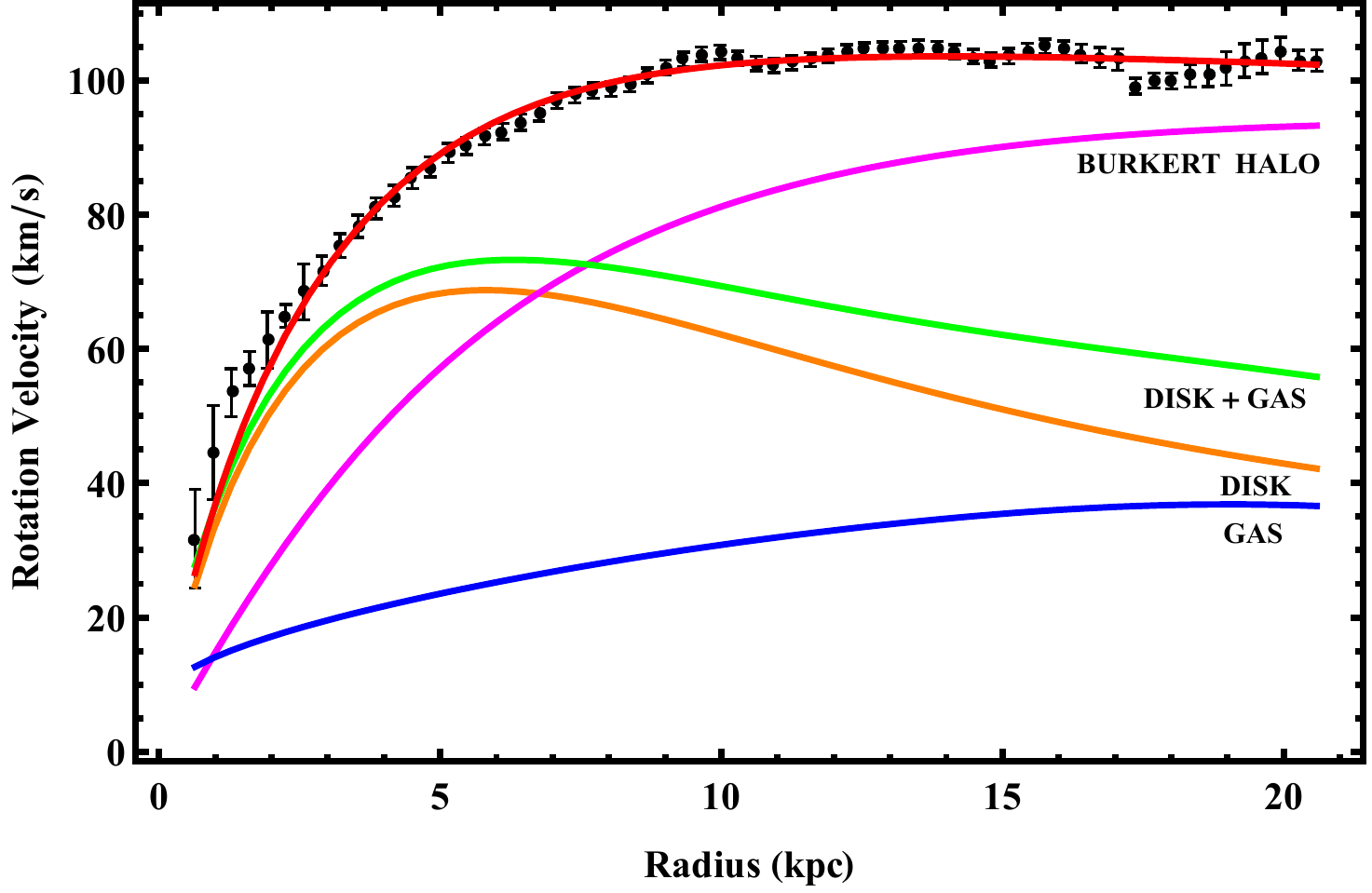}
\caption{A plot of the observed rotation curves for ESO$138$-G$014$ (error bar) and Burkert halo model (red curve) with its separation of disk contribution (orange curve), gas contribution (blue curve), disk and gas contribution (green curve) and halo contribution (magenta curve).}
\label{fig:DiskGasHALO}
\end{figure}

\begin{figure}
\includegraphics[scale=0.6]{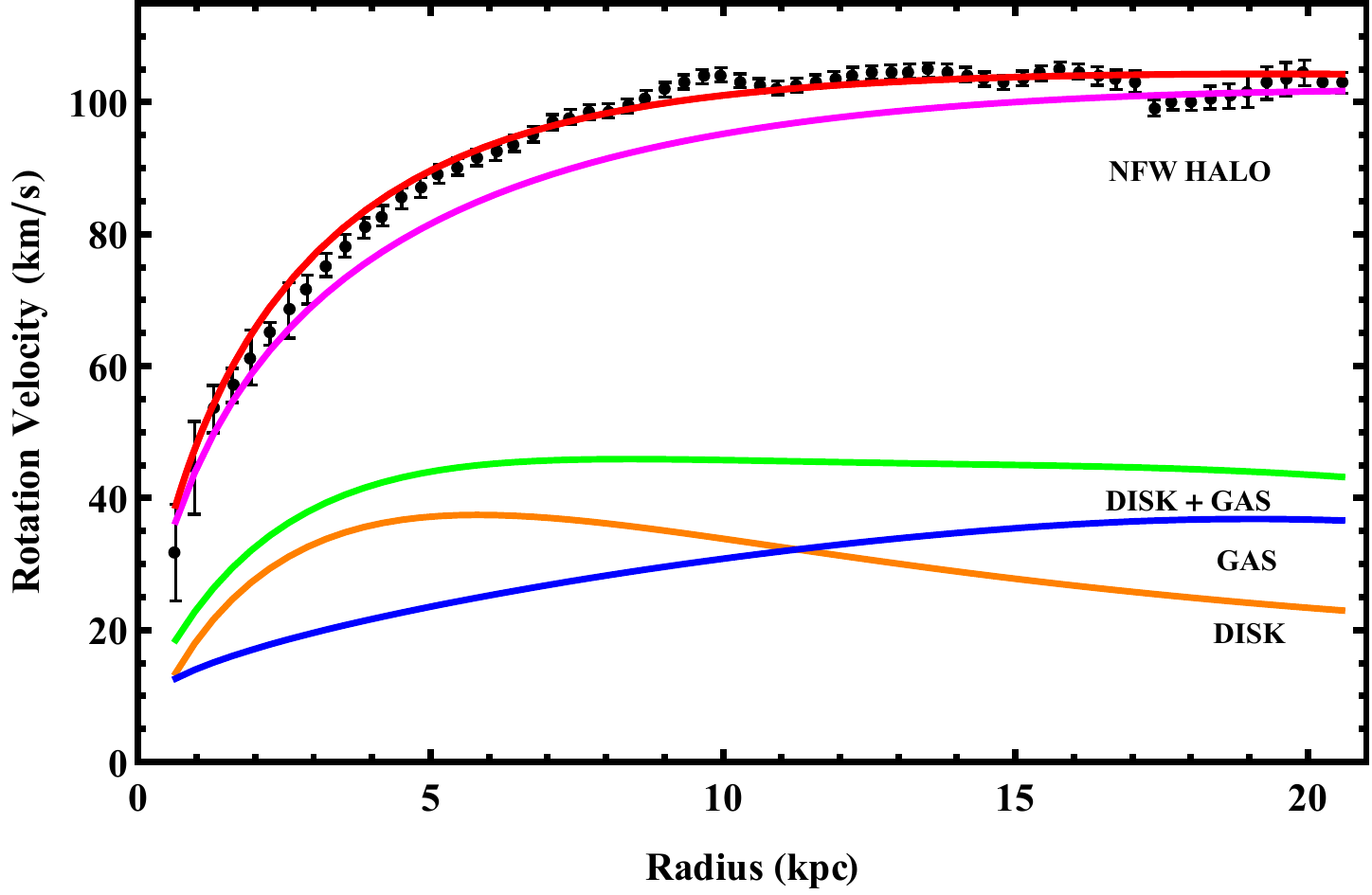}
\caption{A plot of the observed rotation curves for ESO$138$-G$014$ (error bar) and NFW halo model (red curve) with its separation of disk contribution (orange curve), gas contribution (blue curve), disk and gas contribution (green curve) and halo contribution (magenta curve).}
\label{fig:DiskGasHALONFW}
\end{figure}

\begin{figure}
\includegraphics[scale=0.7]{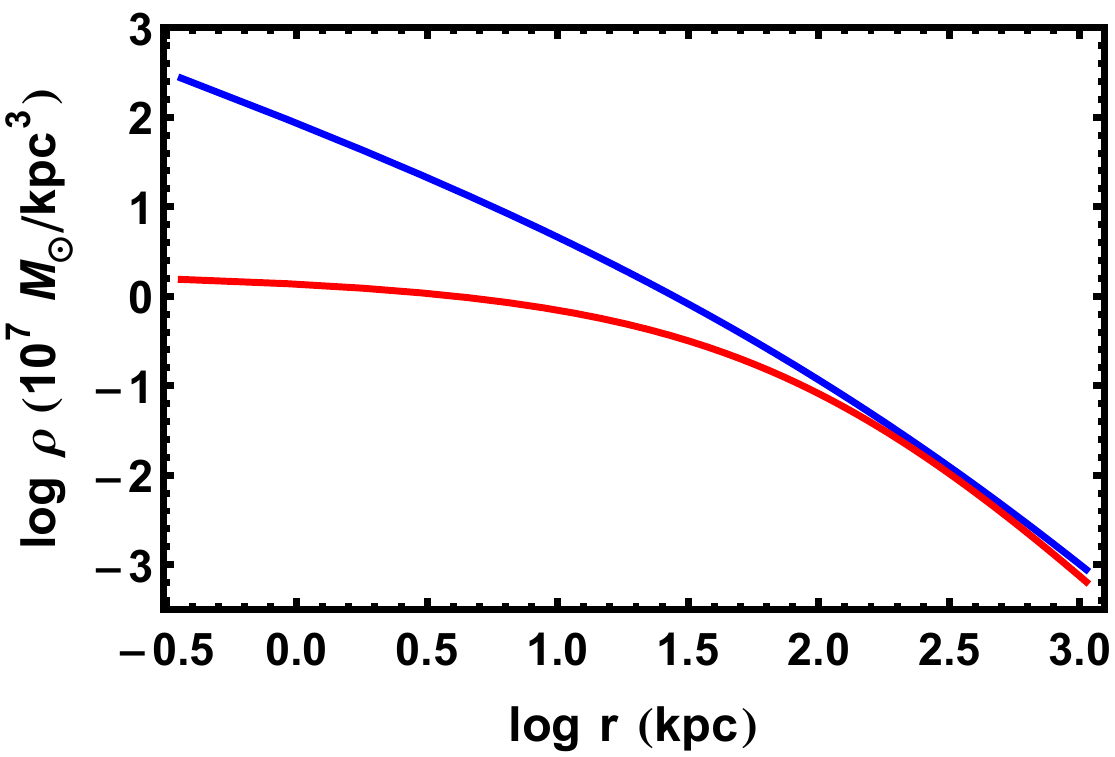}
\caption{The comparison between Burkert density halo profile (red line) and NFW density halo profile (blue line) of ESO$138$-G$014$ galaxy.}
\label{fig:BurkertNFW}
\end{figure}

In this paper, we used nonlinear least square method in Mathematica coding to do the data fitting for two models, i.e Burkert and NFW profiles. The coding also provides statistical descriptions of the fitting to ensure the goodness of fit to the models. The free parameters of the fitting are given in Table~\ref{table1} and the curve fitting results for both models are presented in Figure~\ref{fig:DiskGasHALO} and Figure~\ref{fig:DiskGasHALONFW}. In Table~\ref{table1}, we found that the Burket halo model gives the best fitting of the observed data compared to NFW halo model with $\chi _{red}^{2}=1.4$ and $\chi_{red}^{2}=2.3$, respectively.
\begin{table*}
\caption{The standard error for the models' parameters with reduced chi-square of the fitting.}
\begin{tabular}{ c c c c c r r r }
\hline
Model & $M_{D}$ $[M_{\odot }]$ & $\rho_{0}$ $[M_{\odot }$ kpc$^{-3}]$ & $r_{0}$ [kpc] & $r_{s}$ [kpc] & $\chi _{red}^{2}$\\
\hline
Burkert & $(7.7\pm 0.5)\times 10^{9}$ & $(1.3\pm 0.2)\times 10^{7}$ & $7.5\pm 0.5$ & - & 1.4\\
NFW & $(2.3\pm 2.0)\times 10^{9}$ & $(7.7\pm 4.2)\times 10^{6}$ & - & $10.7\pm 2.8$ & 2.3\\
\hline
\end{tabular}
\label{table1}
\end{table*}
Consequently, up to the last observed point of the rotation curve, we derived the Burkert halo mass from the equation (\ref{Mburkert}) and obtained: $M_{b}\left( 5\right) =4.2\times 10^{10}M_{\odot }$. This is already significantly exceeding the total luminous mass of the entire galaxy. Furthermore, it is in the range of halo mass, $3\times10^{10} M_{\odot}\leq M_{h} \geq 3\times10^{13} M_{\odot}$ that predicted by~\cite{salucci:13}. Halo mass is one of the crucial fundamental physical quantity that characterized the spiral galaxies. We managed to get the expected total mass in agreement with the dynamical mass of the ESO138-G014 galaxy, i.e $M_{dyn}\simeq 5.71\times 10^{10}M_{\odot }$, with 87\% accuracy. Concerning the dark matter distribution, the exact fitting with the Burkert profile shows that the dark halo has a central constant-density core,  $\rho _{b}=4.1\times 10^{5}M_{\odot }$ kpc$^{-3}$ and density profile of this galaxy is shown in Figure~\ref{fig:BurkertNFW}. By applying equation (\ref{NFWdensity}), we can see the dark matter distribution from the NFW density profile (see Figure~\ref{fig:BurkertNFW}). Here we can conclude that we have been successfully constructed the rotation curve model that consisting of the observable components and a cored dark halo. However, we have to take acount the other breadth of this paper, i.e. the alternative of dark matter as to complete our investigation on kinematics of ESO138-G014 galaxy.
\subsubsection{Disk Mass-to-Light Ratio}\label{ML}
Here we investigate the disk mass-to-light ratio of ESO$138$-G$014$ galaxy. Although in practice, the disk mass to light ratio is not precisely known, but we can estimate the disk mass-to-light ratio in this galaxy by the following formula,
\begin{equation}
\bigskip \left( \frac{M}{L}\right) =\frac{M_{D}}{Luminosity}. \label{ML}
\end{equation}
By taking the best value of $M_D$ from the fitting and the total apparent magnitude in $B$ band from the extragalactic database, Hyperleda\footnote{http://leda.univ-lyon1.fr }. The ratio between the disk mass and the luminosity of a galaxy, $L_{B}=5.0\times 10^{9}L_{\odot }$ is $\left( \frac{M}{L}\right) _{B}\sim 1.5$. It is well in agreement with a typical ratio of the disk mass-to-light ratio, of 0.5 - 2.0.

\section{MOND}\label{MOND}
In this section, we will investigate the observed rotation curve in MOND framework. The possibilities of the circular motion within the MOND framework is important to study the validity of this alternative of dark matter. As follow in \cite{gentile:08}, we have
\begin{equation}
V_{MOND}^{2}=V_{bar}^{2}\left( r\right) +V_{bar}^{2}\left( r\right) \left( 
\frac{\sqrt{1+\frac{4a_{0}r}{V_{bar}^{2}\left( r\right) }}-1}{2}\right), \label{Vobs}
\end{equation}
where
\begin{equation}
V_{bar}\left( r\right) =\sqrt{V_{stars}^{2}\left( r\right)
+V_{gas}^{2}\left( r\right) }. \label{Vbar}
\end{equation}
The measured baryon distribution, $V_{stars}(r)$ and $V_{gas}(r)$ are the Newtonian contribution without the bulge contribution to the rotation curve as shown in \cite{milgrom:83}. The second term of equation (\ref{Vobs}) acts like a "pseudo-dark matter halo" and will vanish in the limit $a_{0}\rightarrow 0$. $a_{0}$ is a new fundamental constant, or so-called critical acceleration constant, at which is set below the invalidity of Newtonian gravity. The original proposal by \cite{milgrom:83} suggested, that $a<a_{0}=1.2\times 10^{-8}$ cm s$^{-2}$, could describe the dynamics of galaxies without the dark matter component. Instead, in this paper, we consider the more recently value of $a_{0}=1.35\times 10^{-8}$ cm s$^{-2}$ from the work of \cite{Famaey:07} which is compatible with the result of~\cite{Begemen:91}.

\begin{figure}
\includegraphics[scale=0.55]{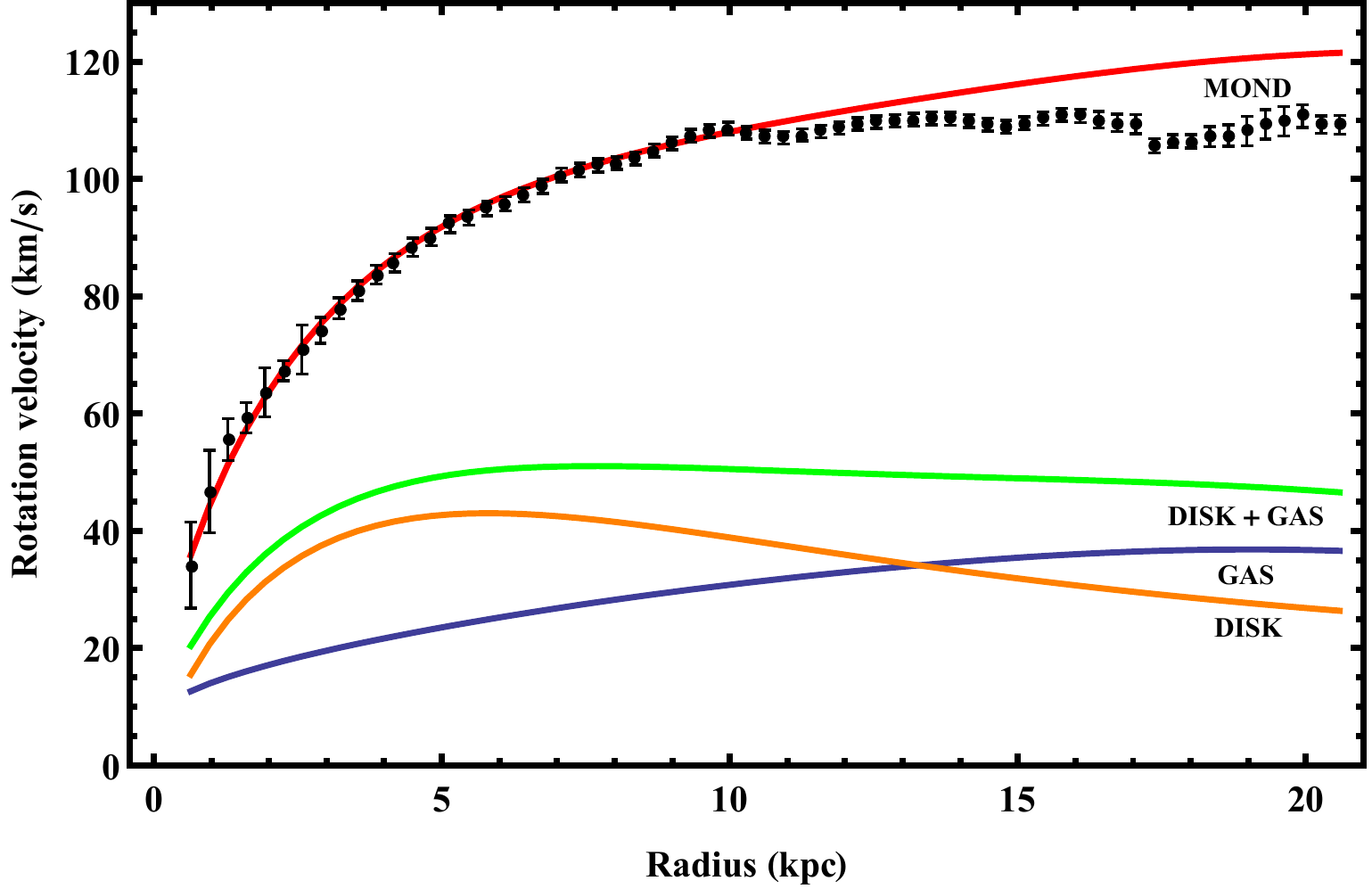}
\caption{The observed rotation curve of ESO$138$-G$014$ (error bar) and MOND prediction with its separation of disk and gas contributions (orange line: disk; blue line: gas estimation from the code with additional factor).}
\label{fig:DiskGasMOND}
\end{figure}

\begin{figure}
\includegraphics[scale=0.55]{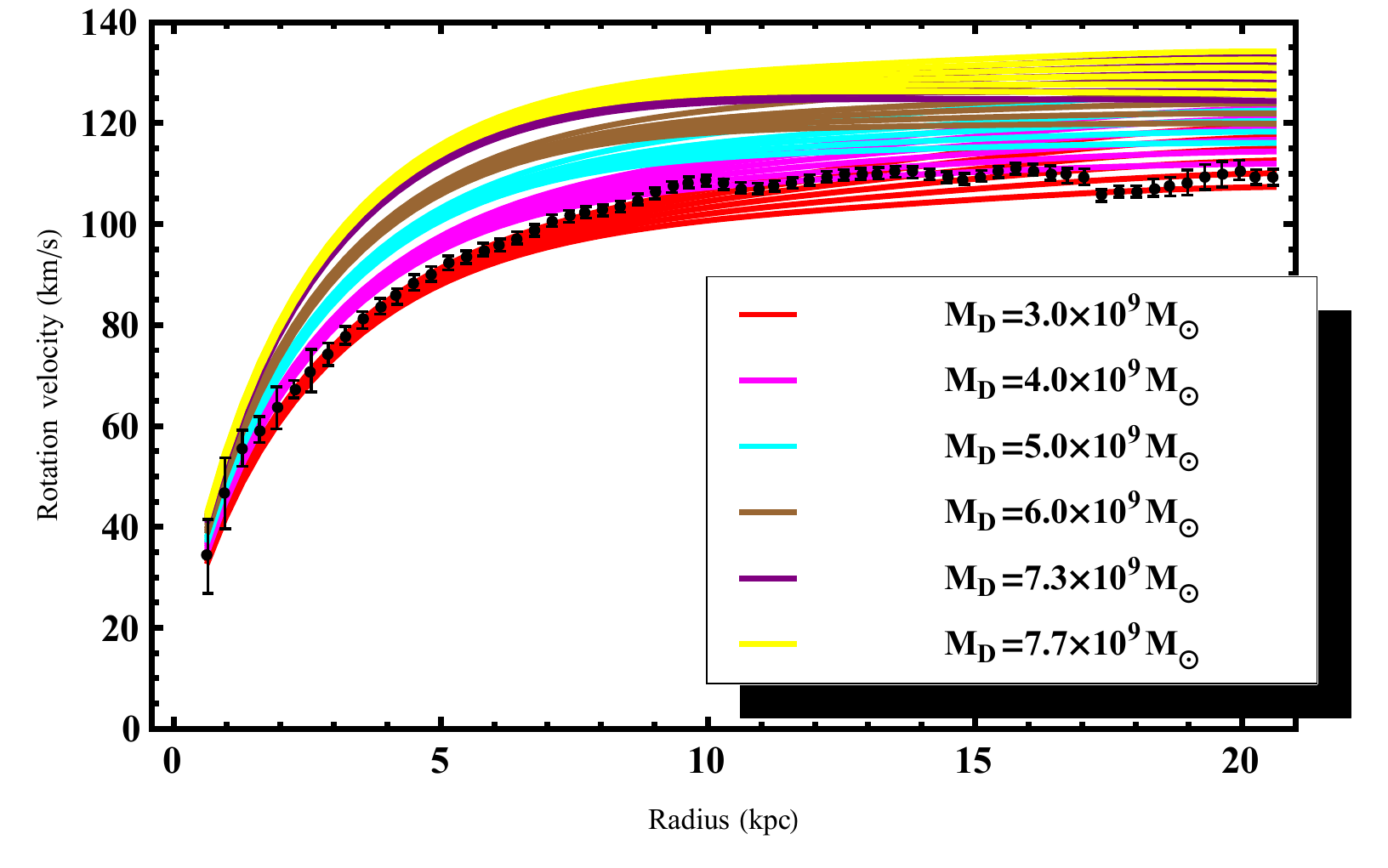}
\caption{The observed rotation curve of ESO$138$-G$014$ (error bar) and MOND prediction with additional factor range $0.5<k<1$ and disk mass range $3.0\times10^{9}M_{\odot}<M_{D}<7.7\times10^{9}M_{\odot}$.}
\label{fig:DiskGasMOND_kMD}
\end{figure}

\begin{figure}
\includegraphics[scale=0.55]{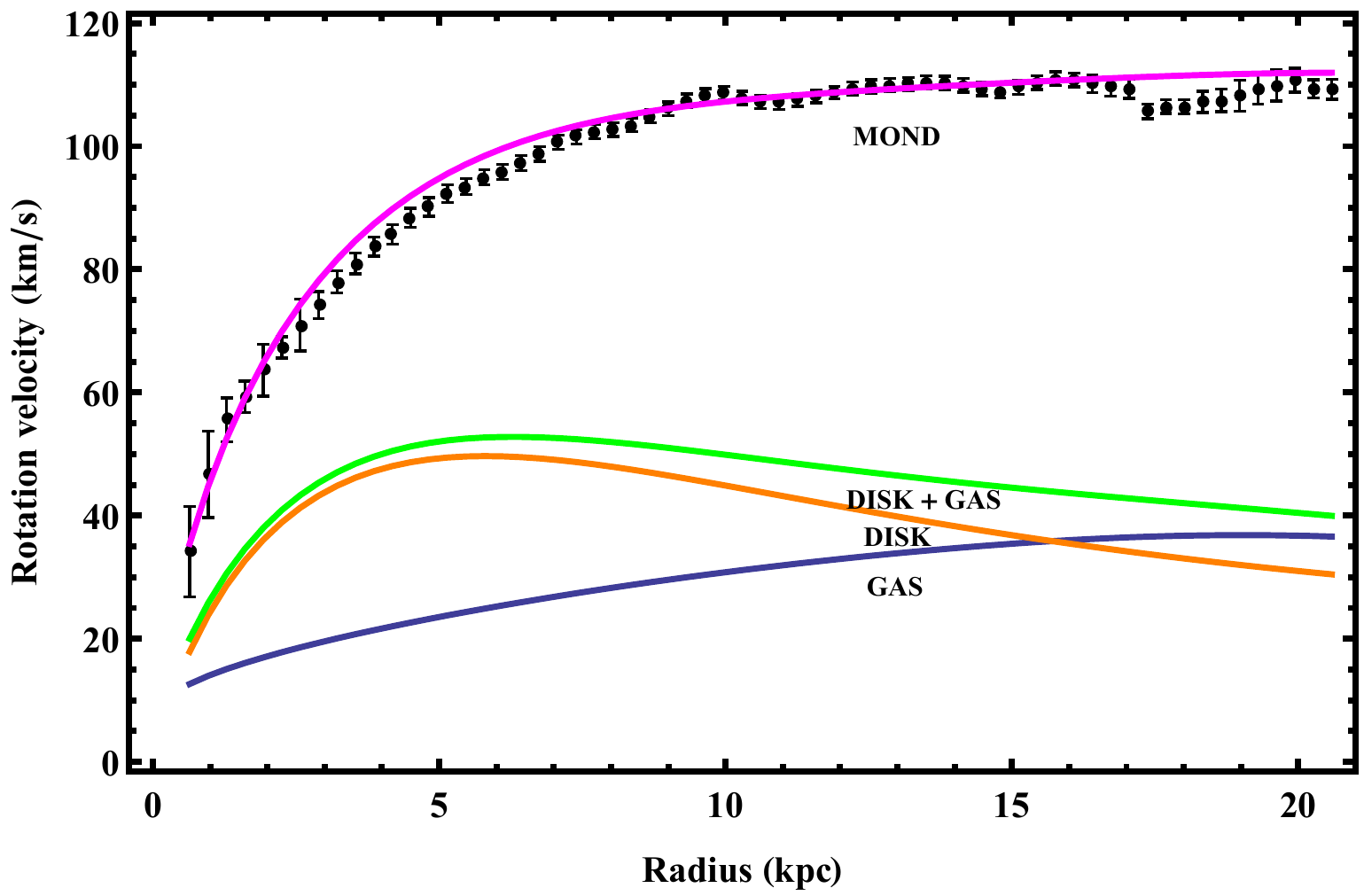}
\caption{The observed rotation curve of ESO$138$-G$014$ (error bar) and the best fit of MOND prediction with additional factor, $k=0.5$ and disk mass, $M_{D}=4.0\times 10^{9}M_{\odot}$.}
\label{fig:MOND_BestFit}
\end{figure}

We obtained the best value of disk mass, $M_{D}$ when we consider equation~(\ref{Vobs}) and (\ref{Vbar}) with $V_{disk}$ (see equation~(\ref{Vdisk})),  $R_{D}\thickapprox\,2.7$ kpc and $V_{gas}$ (see HI mass profile from the Figure~\ref{fig:HImass}), i.e. $M_{D}=3.0\times 10^{9}M_{\odot }$. Then, when we try to add the constant $k=1.1$ to the gas contribution and with the same value of $M_{D}$, the fitting seems not improving but instead worsen the MOND predictions. Here, we can see that the experimental data is not fitted well in MOND profile. The MOND prediction begins to gradually rises up above the experimental data at $12$ kpc from the galactic center (see Figure~\ref{fig:DiskGasMOND}). Similarly, if we set a range of $0.5<k<1$ and increased the value of the disk mass, $M_{D}$ covering the disk mass estimated by rotation curve fitting, the curve fit is unable to reproduce the rotation curve of ESO138-G014 galaxy mostly at all radii. See the reflection of this result in Figure~\ref{fig:DiskGasMOND_kMD}. Figure~\ref{fig:MOND_BestFit} shows the best predicted  MOND result of all which is different from what we have predicted before. The value of $M_{D}$ and $k$ are respectively $4\times10^9 M_{\odot}$ and 0.5. The fully details of the fitting are complied in Table~\ref{table2}. All fits have the values of $\chi_{red}^2 > 1$ manifest the MOND prediction is inappropriate model. If MOND is compatible with the observed data, the $\chi_{red}^2 \approx 1$. Obviously, we have shown that MOND is in disagreement with the observed data.
\begin{table}
\caption{The comparison of the additional factor, $k$ and $M_{D}$ according to the reduced chi-square of the fitting.}
\centering
\begin{tabular}{ c c c c r r r r r r r r r r r r r r r r r r r}
\hline
Model & $M_{D}$, $[M_{\odot }]$ & $k$ & $\chi _{red}^{2}$\\
\hline
MOND 1 & $3.0\times 10^{9}$ & 0.5 & 16.3\\
              & $                            $ & 0.6 & 9.3\\
              & $                            $ & 0.7 & 5.9\\
              & $                            $ & 0.8 & 5.5\\
              & $                            $ & 0.9 & 7.8\\
              & $                            $ & 1.0 & 12.5\\
\hline
\textbf{MOND 2} & \boldmath{$4.0\times 10^{9}$} & \textbf{0.5} & \textbf{3.2}\\
              & $                            $ & 0.6 & 6.7\\
              & $                            $ & 0.7 & 12.6\\
              & $                            $ & 0.8 & 20.6\\
              & $                            $ & 0.9 & 30.6\\
              & $                            $ & 1.0 & 42.3\\
\hline
MOND 3 & $5.0\times 10^{9}$ & 0.5 & 29.5\\
              & $                            $ & 0.6 & 40.2\\
              & $                            $ & 0.7 & 52.6\\
              & $                            $ & 0.8 & 66.6\\
              & $                            $ & 0.9 & 82.1\\
              & $                            $ & 1.0 & 98.9\\
\hline
MOND 4 & $6.0\times 10^{9}$ & 0.5 & 83.0\\
              & $                            $ & 0.6 & 99.0\\
              & $                            $ & 0.7 & 116.3\\
              & $                            $ & 0.8 & 134.8\\
              & $                            $ & 0.9 & 154.4\\
              & $                            $ & 1.0 & 175.0\\
\hline
MOND 5 & $7.3\times 10^{9}$ & 0.5 & 181.8\\
              & $                            $ & 0.6 & 202.8\\
              & $                            $ & 0.7 & 224.8\\
              & $                            $ & 0.8 & 247.6\\
              & $                            $ & 0.9 & 271.3\\
              & $                            $ & 1.0 & 295.8\\
\hline
MOND 6 & $7.7\times 10^{9}$ & 0.5 & 217.5\\
              & $                            $ & 0.6 & 239.7\\
              & $                            $ & 0.7 & 262.8\\
              & $                            $ & 0.8 & 286.8\\
              & $                            $ & 0.9 & 311.5\\
              & $                            $ & 1.0 & 336.9\\
\hline
\end{tabular}
\label{table2}
\end{table}
\section{Conclusions}\label{conclusions}
This paper have successfully described one of the composition of ESO$138$-G$014$ galaxy is dark matter whereby the fitting is in excellent agreement with the Burkert halo profile. It shows that the dark matter is distributed as constant density core rather than cuspy halo. Thus we have probe the dark matter distribution, one of the critical important nature of dark matter for ESO$138$-G$014$ galaxy. In contrast, as we construct the MOND framework to the ESO$138$-G$014$ galaxy, we found that it is a failure for alternative prediction to fit the observed rotation curve. Whereas it seems likely that the failure of the fitting shows that not all the rotation curves of the galaxies can be described by MOND prediction. Nevertheless, despite of the failure of MOND in ESO$138$-G$014$ galaxy, it is obliged to give a credit to the impressive theory of MOND.

\section*{Acknowledgments}
NH thanks the University Malaya (UM) and the Research Training Fellowship from International School for Advanced Studies (SISSA) that have supported this research during of her short visit in Trieste, Italy. Thanks to the University of Malaya grant (UMRG-RG155/11AFR). Special thanks to P. van der Kruit who provided our with the most useful data of the ESO$138$-G$014$ galaxy. MDL and PS  acknowledge the support of INFN Sez. di Napoli and Trieste (Iniziativa Specifica QCSKY).


\end{document}